\documentclass[epj]{svjour}

\usepackage{graphicx}
\usepackage{fancyhdr}
\usepackage{xspace}
\usepackage{array}
\def\makeheadbox{{%  remove EPJC header
 }}
\newcommand{\gluino}   {$\tilde{g}$\xspace}
\newcommand{\squark}   {$\tilde{q}$\xspace}
\newcommand{\mgluino}  {$m_{\tilde{g}}$\xspace}
\newcommand{\msquark}  {$m_{\tilde{q}}$\xspace}
\newcommand{\met}      {$E_{T}^{\mathrm{miss}}$\xspace}
\newcommand{\meff}     {$M_{\mathrm{eff}}$\xspace}
\newcommand{\mt}       {$M_{T}$\xspace}
\newcommand{\st}       {$S_{T}$\xspace}

\def\mytitle{My title} 
\def\myauthors{My name}  
\def\mytype{My type of session}
\def\mysession{My session}
\def\mytitle{Strategy for early SUSY searches at ATLAS} %Put your title here!
\def\myauthors{Shimpei Yamamoto}    %Put your name here!
\def\mytype{Contributed Talk}    
\def\mysession{Colliders - SUSY Phenomenology}
\begin{document}
\title{\mytitle}
%\subtitle{Do you have a subtitle?\\ If so, write it here}
\author{\myauthors
% \thanks is optional - remove next line if not needed
\thanks{\emph{Email:} shimpei.yamamoto@cern.ch}%
\ (for the ATLAS collaboration)
%\and
%Second author\inst{2}% etc
% \thanks is optional - remove next line if not needed
%\thanks{\emph{Present address:} Insert the address here if needed}%
}                     % Do not remove
%
%\offprints{}          % Insert a name or remove this line
%
\institute{International Center for Elementary Particle Physics, the University of Tokyo\\ 7-3-1 Hongo, Bunkyo-ku, Tokyo 113-0033, Japan}
%
%\date{Received: date / Revised version: date}
% The correct dates will be entered by Springer
\date{}
\abstract{
The CERN Large Hadron Collider (LHC) is scheduled to commence operation in 2008 and inclusive searches for supersymmetry (SUSY) will be one of our primary tasks in the first days of LHC operation.
It is certain that the final state of ``multijets + missing transverse energy'' will provide a superior performance in SUSY searches.
Strategies to understand the instrumental background and to understand the Standard Model (SM) background are still under development and are urgent issues for the coming data.
We describe the strategy for early SUSY searches at the ATLAS experiment using the fist data corresponding to an integrated luminosity up to $1\mathrm{fb}^{-1}$, which comprises much progress in the data-driven technique for the SM background estimations.
\PACS{
      {11.30.Pb}{Supersymmetry}   %\and
      %{PACS-key}{discribing text of that key}
     } % end of PACS codes
} %end of abstract
\maketitle
\section{Introduction} \label{intro}
Supersymmetry (SUSY) imposes a new symmetry between the fermions and bosons~\cite{Wess,Fayet,Nilles,Haber,Barbieri}.
The supersymmetric extension of the Standard Model (SM) makes improvements to phenomenological problems in the physics of elementary particles: it provides a natural solution for the gauge hierarchy problem, if sparticles exist at the TeV scale.
Moreover, the extrapolation of LEP data within the framework of supersymmetric extension yields a precise unification of gauge couplings at a scale of $\sim 10^{16}~\mathrm{GeV}$~\cite{Boer,Ellis}.
Due to these properties, SUSY is one of the most attractive alternatives beyond the SM and has been the subject of many studies in particle physics.
However, up to now, no direct evidence for SUSY has been found.
It is essential to examine the properties of any new states of matter at energy scales close to the threshold for the new phenomena, or in high energy collisions at the TeV energy scale.

The CERN Large Hadron Collider (LHC), a $p$-$p$ collider, is scheduled to operate at the center-of-mass energy of $14~\mathrm{TeV}$ in 2008.  
The ATLAS experiment is one of two general purpose experiments being nearly constructed for the CERN LHC.
The ATLAS detector is designed to have a good sensitivity to the full range of high-$p_{T}$ physics in $p-p$ collisions.
Details regarding the detector and its performance are described in Ref.~\cite{TDR}.
Currently we focus on preparation for the early data in 2008, aiming for early SUSY discoveries.
We describe the strategy for early SUSY searches at the ATLAS experiment using the data from the first year of data taking corresponding an integrated luminosity of $1~\mathrm{fb}^{-1}$.
The results in this paper are obtained using the full detector simulation to understand reconstruction performances, trigger efficiencies and systematic effects of the detector.

\section{Inclusive searches}
Sparticle production of gluinos (\gluino) and squarks (\squark) occurs dominantly via strong interactions and its rate may be expected to be considerably large at the LHC.
Gluino production leads to a large rate for events with multijets via series of cascade decay and the neutral lightest supersymmetric particle (LSP) in the final state which remains stable and undetectable, if $R$-parity is conserved. 
LSP's carry off apparently large missing transverse energy (\met).
We note that there are no third generation partons in the initial state, gluino and squark production rates are fixed by QCD in terms of the gluino and squark masses (\mgluino and \msquark).
Thus, inclusive SUSY searches with the early data rely on excesses of events in the channel of ``multijets + large \met''~\cite{Baer} which is a model-independent feature.

In our searches, the events are classified based on the topology, the number of identified isolated leptons, and we try to cover a broad range of experimental signatures.
The experimental signatures, corresponding SUSY scenarios and their background processes are summarized in Table~\ref{tab:1}.
They cover realistic supersymmetric models of minimal supergravity (mSUGRA)~\cite{Chamseddine,Barbieri2,Ohta,Hall}, anomaly-mediated SUSY breaking (AMSB)~\cite{Randall,Giudice} and gauge-mediated SUSY breaking (GMSB)~\cite{Dine,Dine2}.

\begin{table*}
\caption{Summary of experimental signatures with \met and corresponding SUSY scenarios and SM background processes.}
\label{tab:1} 
\begin{center}
\begin{tabular}{cccc}
%\begin{tabular}{p{3zw}p{5zw}p{9zw}p{5zw}}
\hline\hline\noalign{\smallskip}
%$N_{\mathrm{jet}}$ & additional signature & covered scenario & background \\
jet multiplicity & additional signature & covered scenario & background \\
\noalign{\smallskip}\hline\noalign{\smallskip}
$\ge 4$  & no lepton & mSUGRA, AMSB, split SUSY, heavy squark & QCD, $t\bar{t}$, $W/Z$ \\
         & single lepton ($e$,$\mu$) & mSUGRA, AMSB, split SUSY, heavy squark & $t\bar{t}$, $W$ \\
         & dilepton ($e$,$\mu$) & mSUGRA, AMSB, GMSB & $t\bar{t}$ \\
         & ditau & GMSB, large $\tan\beta$ & $t\bar{t}$, $W$ \\
         & $\gamma\gamma$ & GMSB & --- \\
$\sim 2$ & --- & light squark & $Z$ \\
\noalign{\smallskip}\hline\hline
\end{tabular}
\end{center}
\vspace*{1cm} 
\end{table*}

\subsection{Event selection}
Considering the features of sparticle production and its decay, the signal candidate events are selected by requiring:
\begin{itemize}
  \item $N_{\mathrm{jet}} \ge 4$,
  \item $p_{T}^\mathrm{J1} > 100~\mathrm{GeV/c} \ \& \ p_{T}^\mathrm{J4} > 50~\mathrm{GeV/c}$,
  \item $S_{T} > 0.2$,
  \item $E_{T}^{\mathrm{miss}} > 100~\mathrm{GeV} \ \& \ E_{T}^{\mathrm{miss}} > 0.2 \times M_{\mathrm{eff}}$,
\end{itemize}
where $N_{\mathrm{jet}}$, $p_{T}^\mathrm{J1(4)}$, \st and \meff are the number of jets , the transverse momentum of first (fourth) leading jet, the transverse sphericity and the effective mass, respectively.
The effective mass is formulated as $ M_{\mathrm{eff}} = \sum_{i=0}^{i\le4} p_{T}^{i} + E_{T}^{\mathrm{miss}}$, where $p_{T}^{i}$ is the transverse momentum of $i$-th leading jet.
In addition, for the single-lepton signature, the events are selected by requiring one isolated lepton with $p_{T}$ larger than $20~\mathrm{GeV}$ and the transverse mass (\mt) should be larger than $100~\mathrm{GeV}$.
The selection cuts described here are based on the definition given in Ref.~\cite{TDR}, but will be optimized.

\begin{figure}[htbp]
  \begin{center}
\includegraphics[width=0.42\textwidth,angle=0]{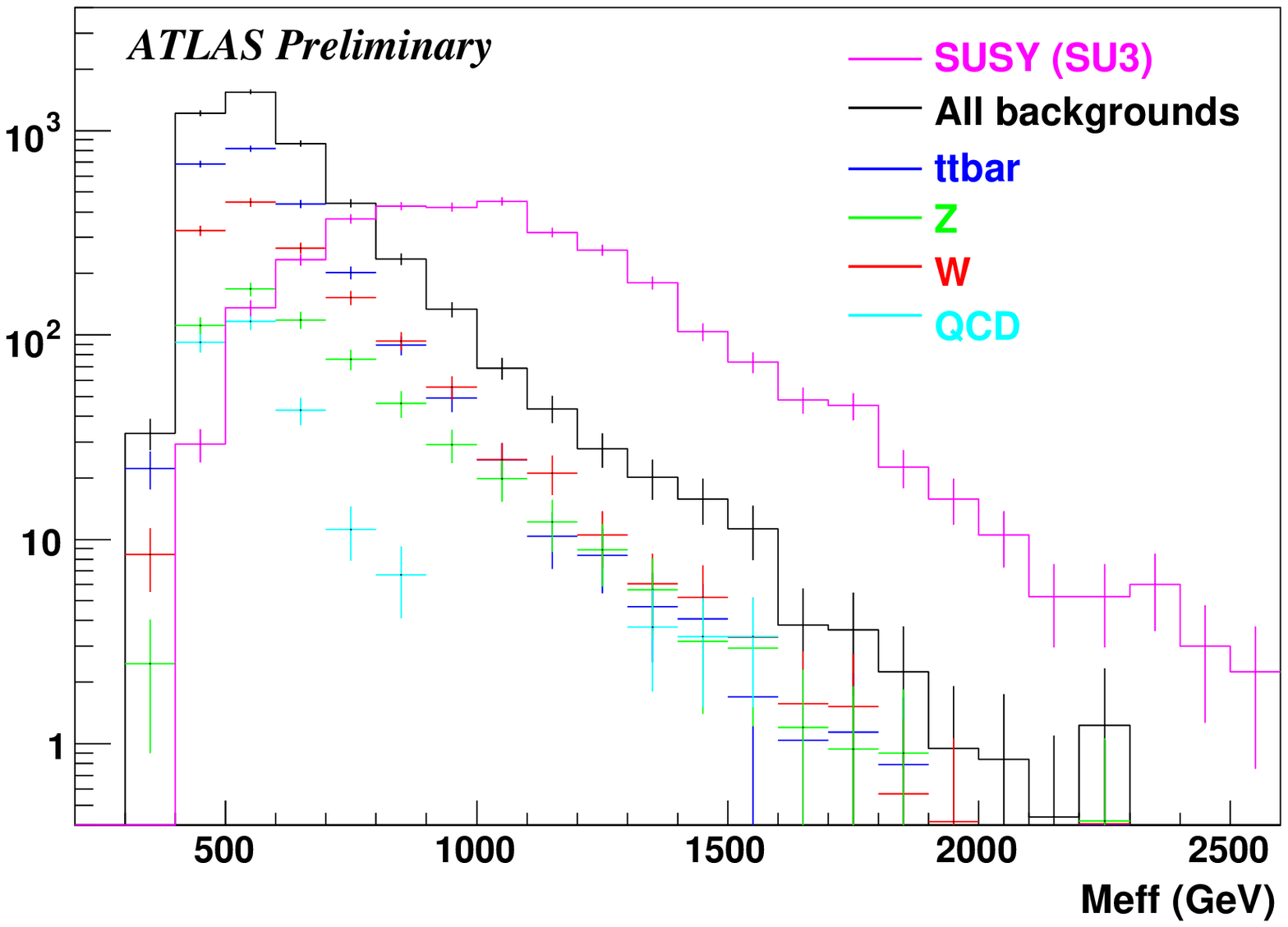}
\includegraphics[width=0.42\textwidth,angle=0]{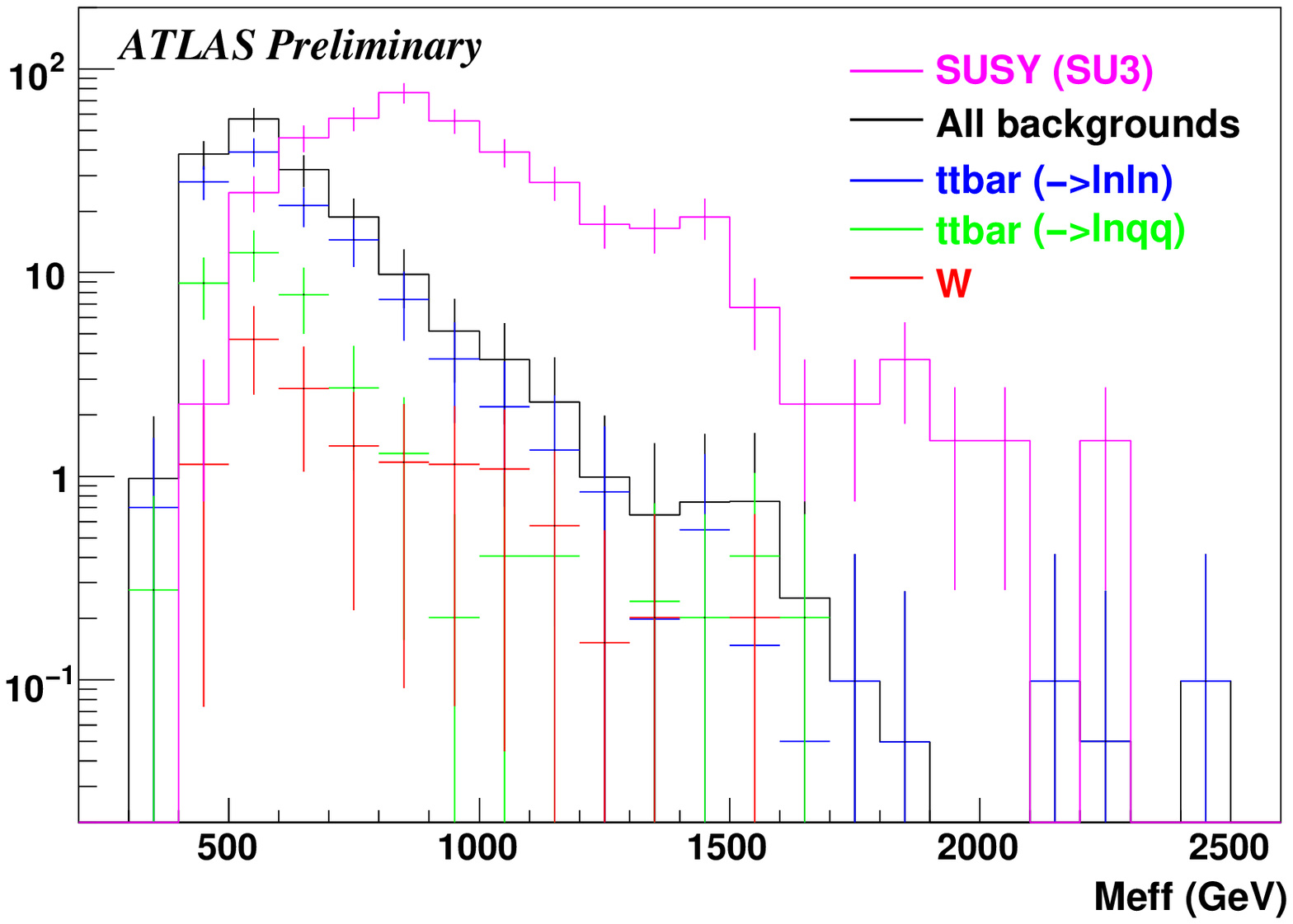}
\caption{The \meff distributions for no-lepton (upper) and single-lepton (bottom) signatures.
The histograms are normalized to the integrated luminosity of $1\mathrm{fb}^{-1}$.
}
\label{fig:meff}
  \end{center}
\end{figure}

Finally, we look for SUSY events with large \meff at which the signal exceeds the SM backgrounds.
The value of \meff also provides a first estimate of the sparticle masses.
Fig.~\ref{fig:meff} shows the \meff distributions for no and single-lepton signatures.
For the SUSY signal, we set a bench mark point in the bulk region, where $m_{0}=100~\mathrm{GeV}$, $m_{1/2}=300~\mathrm{GeV}$, $A_{0}=300~\mathrm{GeV}$, $\tan\beta=6$ and $\mu>0$, referred to SU3.
If SUSY exists we expect there is an excess over the SM background expectation in the distribution for each event topology with an integrated luminosity up to $1\mathrm{fb}^1$.

\subsection{Discovery reach}
\begin{figure}[htbp]
  \begin{center}
\includegraphics[width=0.42\textwidth,angle=0]{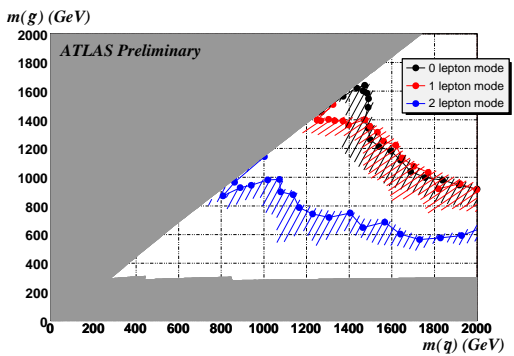}
\caption{$5\sigma$-discovery reaches in the $m_{\tilde{g}}-m_{\title{q}}$ space for each event topology.
Generator-level uncertainties of 100\% are shown by the hatched regions.
The results are obtained on the assumption of $tan\beta=10$, $A_{0}=0$ and $\mu>0$.
}
\label{fig:discovery}
  \end{center}
\end{figure}

Figure~\ref{fig:discovery} shows $5-\sigma$ discovery reach in the \mgluino-\msquark space for each event topology in the mSUGRA model.
We see that even for sparticle mass as heavy as $\sim 1~\mathrm{TeV}$, discoveries are expected for an integrated luminosity of $1\mathrm{fb}^{-1}$.
The discovery reach also shows good stability against the values of $\tan\beta$.

\section{{\it In-situ} measurements for the \met scale and resolution}
As stated above, the \met is the discriminating signature for the SUSY searches, but it is also a complex object: apart from undetected neutral particles, it will comprise contributions from beam halo, cosmic-ray muons and instrumental effects such as noise, hot or dead channels or cracks of the detector.
The mismeasurements or inefficiencies for jets and muons also contribute to the \met; these make up the fake \met.
A precise understanding of these contributions and their reduction are crucial, especially in the tails of the distribution.

\begin{figure}[htbp]
  \begin{center}
\includegraphics[width=0.4\textwidth,angle=0]{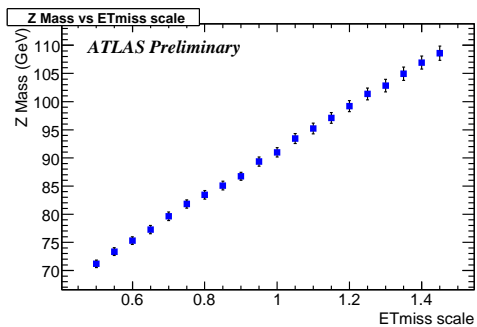}
\caption{The $Z \to \tau\bar{\tau}$ invariant mass as a function of \met scale.}
\label{fig:etmissscale_ztautau}
\end{center}
\end{figure}

For the calibration and commissioning of \met, we use the transverse mass distributions of $W \to l\nu$ and the invariant mass of $Z \to \tau\bar{\tau}$ where $\tau$'s decay into a lepton and hadrons.
For the $Z \to \tau\bar{\tau}$, as an example, the invariant mass is reconstructed using the collinear approximation; it is sensitive to the \met scale and resolution.
Fig.~\ref{fig:etmissscale_ztautau} shows the $Z \to \tau\bar{\tau}$ invariant mass peak as a function of the \met scale.
A variation of 10\% in the \met scale results in a 3\% shift on the $Z$ mass scale.
The $Z$ mass can be reconstructed with an error of 1\% using an accumulated data of $100~\mathrm{pb}^{-1}$.
Considering the uncertainty of 1\% on the absolute energy scale of the calorimeters, we can evaluate the \met scale with an accuracy of $\sim 4\%$ accordingly.

The fake \met originating from the instrumental effects of the detector have also been studied.
All the results for the commissioning are given in Ref.~\cite{CSCnote}.

\section{Data-driven approaches for background estimations}
Even if we get early indications of SUSY, we should justify ``beyond SM'' signatures with full understandings of generator-level uncertainties and instrumental effects of the detector because they affect the normalization and shapes of predicted backgrounds.
However, these uncertainties are hard to estimate in the early stage of the experiment, and they are also expected to be large.
Thus, in the early SUSY searches, we take data-driven approaches for the SM background estimation.

\subsection{$Z/W$ boson production background}
The process of $Z\to \nu\bar{\nu}$ in association with multijets will give rise to final states with large \met and could be a dominant background for the no-lepton signature.
For this background contamination, the expectation is derived from the MC distribution of $Z \to \nu\bar{\nu}$ with the normalization determined by the $Z \to l\bar{l}$ data, where $l$ is $e$ or $\mu$.
%The normalization factor is obtained by comparing the $Z \to \nu\bar{\nu}$ data with the MC events where there are less background in the final state.
We can better measure the $Z \to l\bar{l}$ yield thanks to small backgrounds in the final state.
%We then apply it to the $Z \to \nu\bar{\nu}$ MC events.
%The resultant \meff plot is shown in Fig.\ref{fig:Znunu}.
%This normalization factor can be also applied to the $W \to l\nu$ background due to the same production mechanism at the LHC.
This normalization factor can be also applied to the $W \to l\nu$ background.
Differences due to different masses and production mechanisms can be taken from theory.
%Both estimations work sufficiently.

\subsection{QCD multijet background}
\begin{figure}[htbp]
  \begin{center}
\includegraphics[width=0.25\textwidth,angle=0]{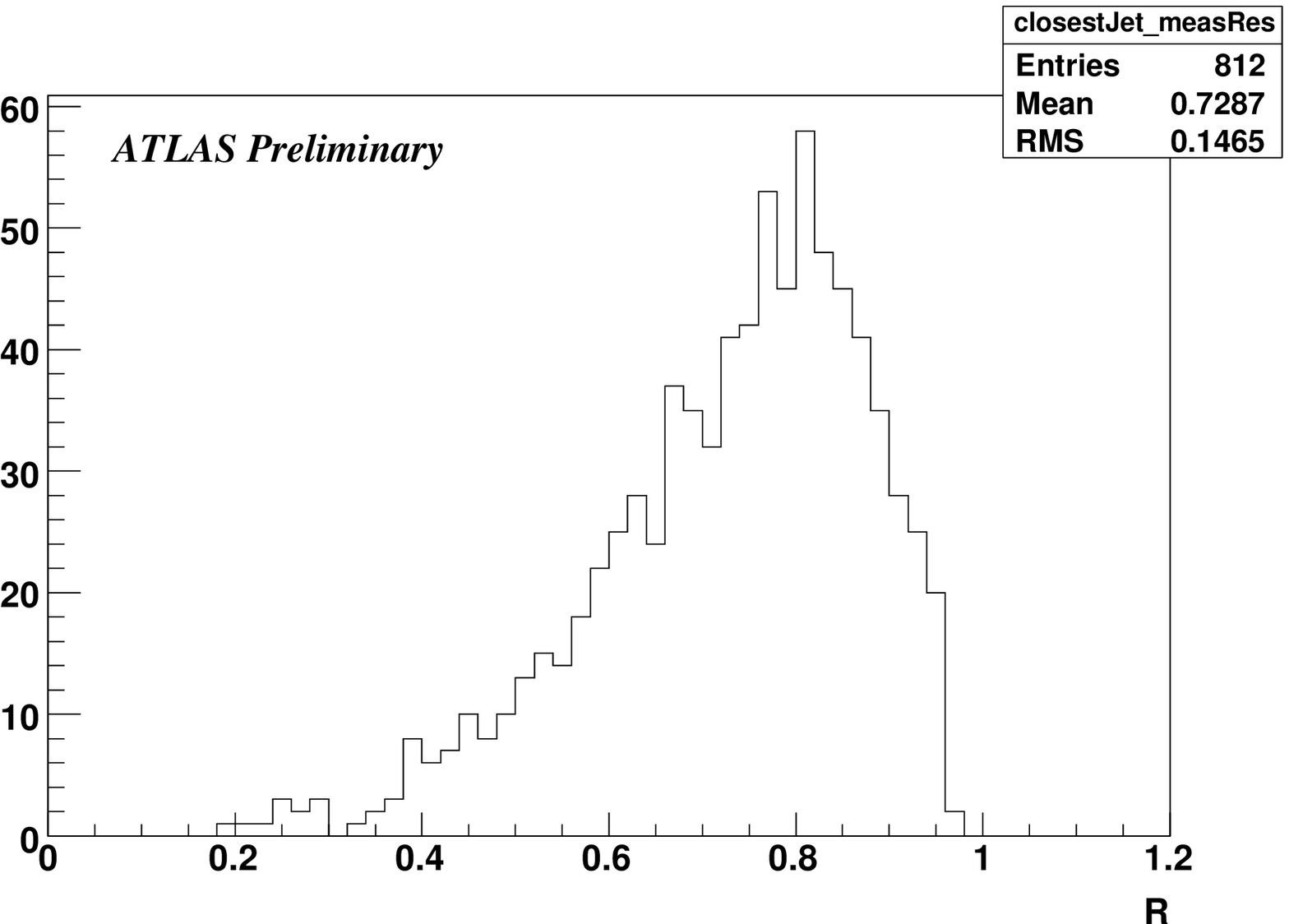}
\includegraphics[width=0.23\textwidth,angle=0]{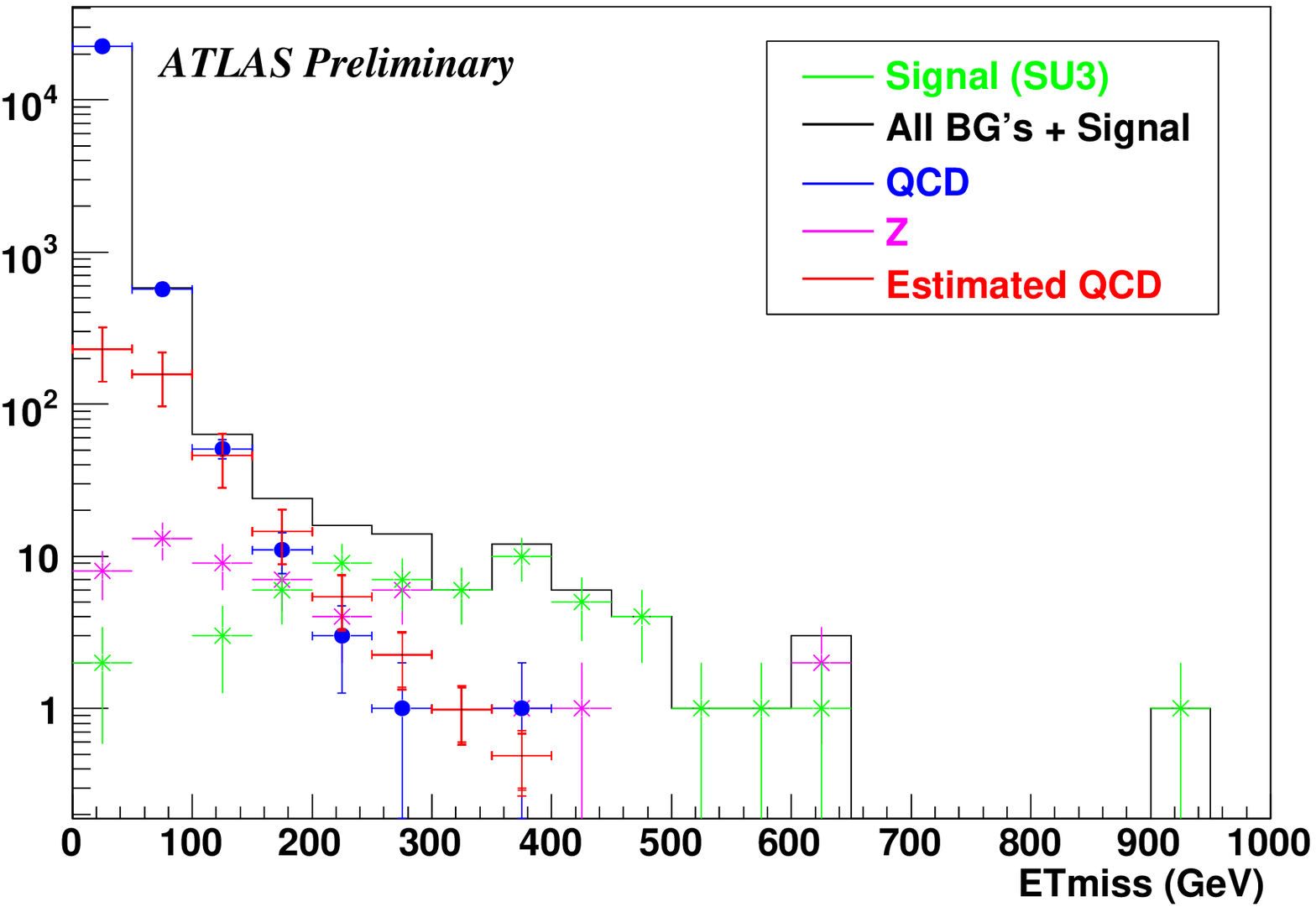}
\caption{The jet fluctuating function (left) and the \met distribution (right) for the no-lepton signature.
The histograms are normalized to an integrated luminosity of $22~\mathrm{pb}^{-1}$.
}
\label{fig:QCDbg}
  \end{center}
\end{figure}

The QCD multijet production could be one of the most dominant SM background sources due to its large cross section.
In QCD events, neutrinos via leptonic decays and the mismeasurements of jets contribute to the tails of \met distribution.
For the QCD background contamination, the estimation is derived from the multijet data with a function representing the fluctuations of measured jet energies. 
The fluctuation of jet energy is evaluated using events of
\begin{itemize}
  \item $E_{T}^{\mathrm{miss}} > 60~\mathrm{GeV}$,
  \item $\Delta \phi(E_{T}^{\mathrm{miss}}, \mathrm{jet}) < 0.1$,
\end{itemize}
where $\Delta \phi(E_{T}^{\mathrm{miss}}, \mathrm{jet})$ is the $\phi$-angle between the missing transverse energy and an isolated jet in radian.
We suppose that \met is originated from the fluctuating jet close to the \met direction, and the $p_{T}$ of initial jet is estimated to be the vectorial sum of $p_{T}^{\mathrm{jet}}$ and \met.
Then we can obtain the fluctuating function of jet energies ($R \equiv 1-\mathbf{p}_{T}^{\mathrm{jet}}\cdot(\mathbf{p}_{T}^{\mathrm{jet}}+\mathbf{E}_{T}^{\mathrm{miss}})/|\mathbf{p}_{T}^{\mathrm{jet}}+\mathbf{E}_{T}^{\mathrm{miss}}|^{2}$).
The jets in QCD multijet events with small \met are smeared according to the jet fluctuating function, which result in fluctuations of \met.
Fig.~\ref{fig:QCDbg} shows the jet fluctuating function and the \met distribution with the superimposed QCD background estimation.
We can obtain a fair description of the QCD background with this method, especially in the tails of the distribution.

\subsection{\mt discrimination method}
Another advanced method to estimate the SM backgrounds, especially for single-lepton signature,  is to use \mt which shows a discriminating power between SUSY and SM backgrounds but are less dependent on \met.
At first, we select events of $M_{T} < 100~\mathrm{GeV}$ to enhance the backgrounds (background control sample) and suppose the \met distribution of these events represents that of all the SM backgrounds.
Then, the overall normalization of SM backgrounds is determined by comparing the number of events with $M_{T}<100~\mathrm{GeV}$ and that with $M_{T} \ge 100~\mathrm{GeV}$ in the low \met region below $200~\mathrm{GeV}$.

\begin{figure}[htbp]
  \begin{center}
    \includegraphics[width=0.27\textwidth,angle=0,trim=5mm 0mm 10mm 3mm]{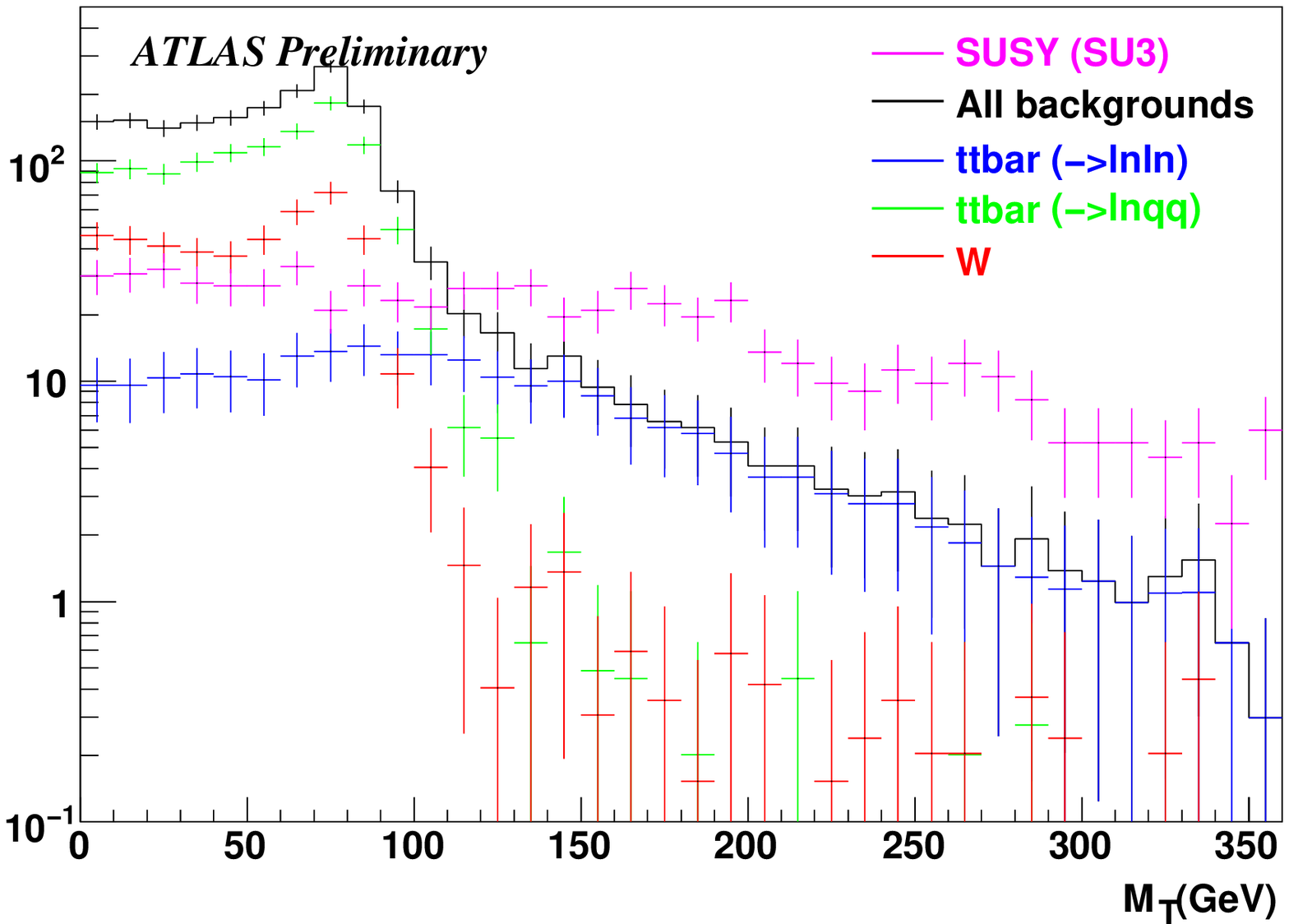}
    \includegraphics[width=0.2\textwidth,angle=0,trim=10mm 0mm 5mm 0mm]{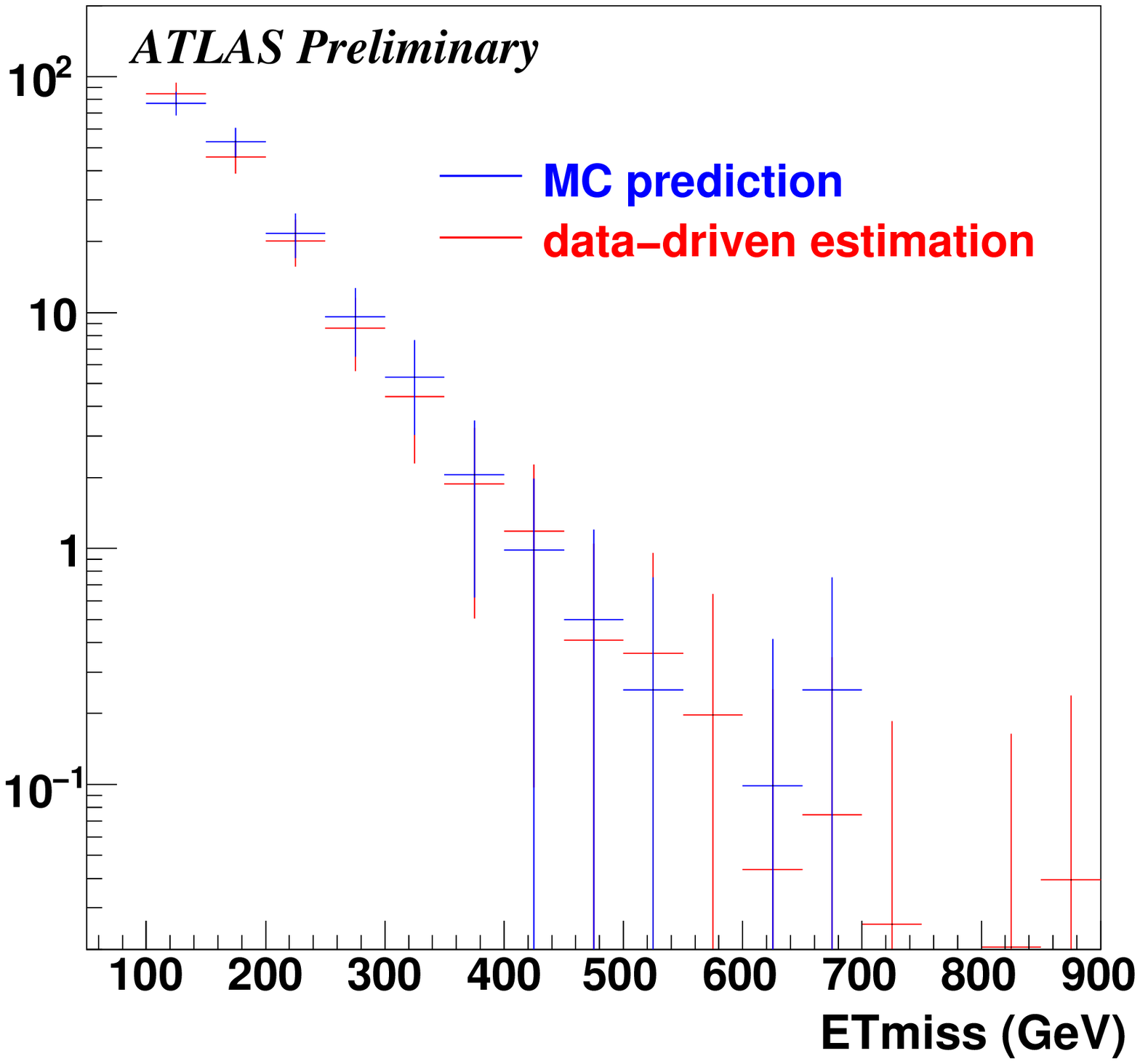}
    \caption{The \mt distribution of single-lepton candidate events (left) and the \meff distribution of SM backgrounds (right).
    The histograms are normalized to an integrated luminosity of $1 \mathrm{fb}^{-1}$.
    }
    \label{fig:MT}
  \end{center}
\end{figure}

The \mt distribution and the resultant \meff are shown in Fig.~\ref{fig:MT}.
The estimated number of SM backgrounds with large \meff ($>800~\mathrm{GeV}$) and the MC prediction are $22.0 \pm 0.9$ and $24.8 \pm 1.6$, respectively.
They show a good agreement, however, there could be a contamination of signal events in the background control sample, which results in the overestimation of backgrounds by a factor of $\sim 2$.
%We are making a progress in this method for the reduction of signal contamination and the result will be described in Ref.~\cite{CSCnote}.
We will make further investigations to improve this method and to reduce the signal contamination in the normalization region.
Results will be reported in Ref.~\cite{CSCnote}.

\section{Conclusions}
We described the strategies for early SUSY searches at the ATLAS experiment.
Using the inclusive signature of multijets plus large \met, we can make an early discovery of SUSY.
Our searches are less model-dependent but sensitive to models with squark and gluino masses up to $1~\mathrm{TeV}$ with an integrated luminosity of $1\mathrm{fb}^{-1}$.
However, the precise understanding of the SM backgrounds is crucial for the early discovery.
To settle this issue, we have been making much progress in the {\it in-situ} calibration of \met and the data-driven technique for the SM background estimations.
The opening of a new era of physics beyond the SM is coming with the LHC.

\section*{Acknowledgments}
The author wishes to thank SUSY, Jet/Etmiss and many other working groups of the ATLAS collaboration.
The work described here relied on their joint efforts.

%
% BibTeX users please use
% \bibliographystyle{}
% \bibliography{}
%
% Non-BibTeX users please use

\end{document}